\documentclass[aps,prb,twocolumn,groupedaddress,showpacs]{revtex4}

\usepackage{graphicx}
\usepackage{amsmath}

\begin{document}

\title{Thermodynamic transitions in inhomogeneous cuprate superconductors}

\author{J.W. Loram$^{1}$ and J.L. Tallon$^{2,3}$}

\affiliation{$^1$Cavendish Laboratory, Cambridge University,
Cambridge CB3 0HE, England.}

 \affiliation{$^2$MacDiarmid Institute, Industrial Research Ltd., P.O. Box
31310, Lower Hutt, New Zealand.}

\affiliation{$^3$Victoria University, P.O. Box 600, Wellington,
New Zealand.}

\date{\today}

\begin{abstract}
Scanning tunneling spectroscopy studies on
Bi$_2$Sr$_2$CaCu$_2$O$_{8+\delta}$ suggest the presence of
electronic inhomogeneity with a large spatial variation in gap
size. Andersen {\it et al} have modelled this variation by
assuming a spatially-varying pairing interaction. We show that
their calculated specific heat is incompatible with the
experimental data which exhibit narrow transitions. This calls
into question the now-common assumption of gap and pairing
inhomogeneity
\end{abstract}


\maketitle

In a recent paper Andersen and coworkers\cite{Andersen} attempt to
model the spectroscopic inhomogeneity inferred from scanning
tunneling spectroscopy (STS) studies on
Bi$_2$Sr$_2$CaCu$_2$O$_{8+\delta}$ (Bi-2212)\cite{Pan,Lang}. These
STS studies suggest a $\sim\pm25\%$ variation in both the gap
magnitude and the local density of states (LDOS), on a length
scale of the order of one or two coherence lengths, $\xi_0$. The
locations with large spectral gaps lack well-developed coherence
peaks and, based on the absence of Ni resonances there, they are
suggested to be non-superconducting\cite{Lang}. These conclusions
have become highly influential.

To model these effects Andersen {\it et al.}\cite{Andersen} adopt
a spatially-inhomogeneous pairing interaction in a $d$-wave BCS
pairing model and solve self-consistently for the local gap
magnitude in order to map the spatially varying interaction onto
the observed variation in gap magnitude. They then compute the
specific heat and show that the breadth of the anomaly is similar
to the spread in local gap magnitude. This breadth, they claim, is
``comparable to experiments on this material", and conclude that
``substantial nanoscale electronic inhomogeneity is characteristic
of the bulk BSCCO system".

However our data\cite{Loram} (see Fig. 1), which they use for
their comparison, points to the very opposite conclusion. They
erroneously equate the extended fluctuation region above $T_c$
with broadening of their mean-field (MF) transition over a 40K
range. Strong fluctuations are an intrinsic property of this
highly-anisotropic material\cite{Junod} but are not included in
their calculations. In fact it is the narrow region of strong
negative curvature close to each peak (see arrows in Fig. 1) that
reveals the true extent of extrinsic broadening. Such a strong
$T$-dependence over a very restricted $T$-range would not be
possible in a material with a broad distribution of $T_c$ values.

Moreover, the locations where their gap is a maximum correspond to
the maximal local pairing interaction i.e. where SC is strongest.
The STS studies\cite{Lang} show the opposite: SC is weakest and
perhaps absent at the points where the supposed gap is maximal.
Their model also implies that there are local regions where SC
persists well above $T_c$, as much as 30K for Bi-2212. Such
regions are not observed in STS\cite{Renner,Yazdani}.

Here we consider the experimental specific heat data in more
detail and show that the transitions are not strongly broadened as
suggested by Andersen {\it et al.}\cite{Andersen}, thus
potentially reversing their inference of inhomogeneous pairing in
the bulk material. This concurs with several other studies which
imply the absence of gross inhomogeneity in cuprate
superconductors\cite{Bobroff,Storey}.

The specific heat near $T_c$ consists of a MF step at $T_c$ and a
(nearly symmetrical) fluctuation contribution above and below
$T_c$\cite{Junod}. More generally, $T_c$ may be broadened out into
a distribution of $T_c$ values. The separate contributions of
fluctuations and transition broadening may seem similar well away
from the mean $T_c$, but nearby they are quite distinctive and
easily separated. Andersen et al.\cite{Andersen} fail to recognize
this distinction.

Let us consider first fluctuations where there is a sharply
defined $T_c$. The fluctuation specific heat should diverge at
$T_c$ but is cut-off due to the inhomogeneity length scale. For
Bi-2212 we have previously analyzed the fluctuation
contribution\cite{Loram} and deduced transition half-widths as
small as $\Delta T_c/T_c \sim 0.014$, consistent with an
inhomogeneity length scale as large as 16$\xi_0$, much greater
than $\xi_0$. The cutoff is reflected in the narrow region of
negative curvature between the inflexion points in the specific
heat coefficient, $\gamma(T)$, just above and below $T_c$ as shown
by the arrows in Fig. 1(a).

\begin{figure}
\centerline{\includegraphics*[width=85mm]{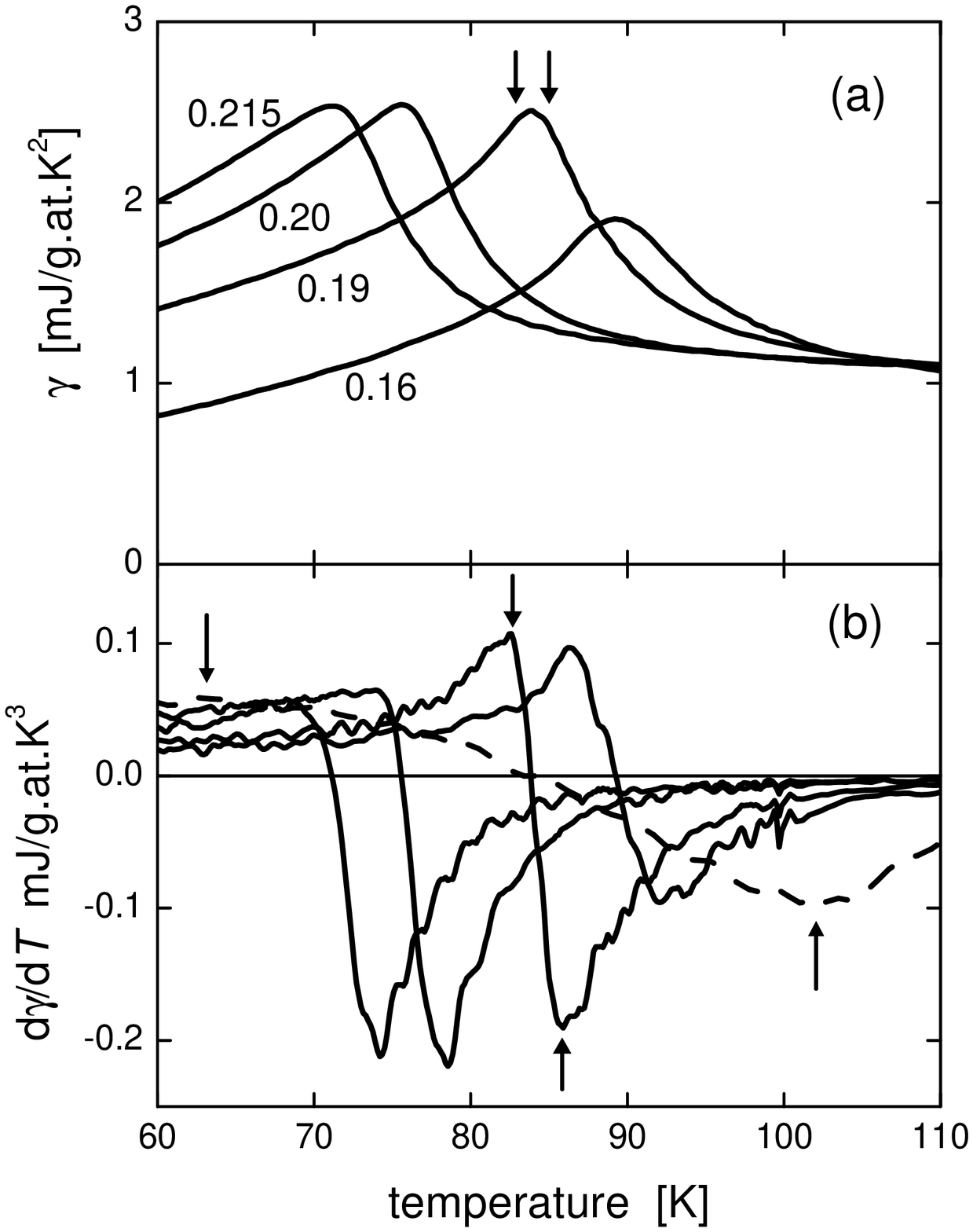}}
\caption{\small (a) The specific heat coefficient, $\gamma$, for
Bi-2212 with $p$= 0.16, 0.19, 0.20 and 0.215, respectively. (b)
The derivative $\partial\gamma/\partial T$. The dashed curve is
$\partial\gamma/\partial T$ from the Andersen {\it et al.}
calculation. Arrows indicate inflexion points in $\gamma(T)$.}
\end{figure}

In Fig. 1(b) we show the derivative $\partial\gamma /\partial T$
from our data and compare it with that from Andersen's model
calculation. The data curves correspond to doping states of $p$ =
0.16, 0.19, 0.20 and 0.215. The inflexion points are located at
the maxima and minima below and above $T_c$, and between them
$\partial\gamma/\partial T$ changes sign. For $p$ = 0.19 the
inflexion points are just 3.3K apart. For the model calculation
they are up to 40K apart (dashed curve and arrows), just as would
be expected for a $\pm25\%$ spread of gap values. This directly
illustrates that we observe rather sharp thermodynamic transitions
in Bi-2212.

To show that the transition is indeed narrow we plot in Fig. 2 the
field dependence of $\gamma$ for a Bi-2212 sample with doping $p
\approx 0.21$. Firstly it is clear from this plot that $T_c(H=0)$
is  close to the peak (as expected if the MF step is small). The
narrow peak is progressively suppressed and broadened by the
field, with a marked effect even for fields as low as 0.3T. The
vortex separation ~ $\sqrt{\phi_0/H} \sim$ 45
nm/$\sqrt{H(\text{Tesla})}$, which acts as an inhomogeneity length
scale, is very large at such low fields and the sensitivity of the
transition to a field as low as 0.3T supports our conclusion that
the order parameter is rather homogeneous. Calculated transitions
based on an inhomogeneity length scale of ~1.6 - 2.5 nm would be
totally insensitive to such low fields.

\begin{figure}
\centerline{\includegraphics*[width=85mm]{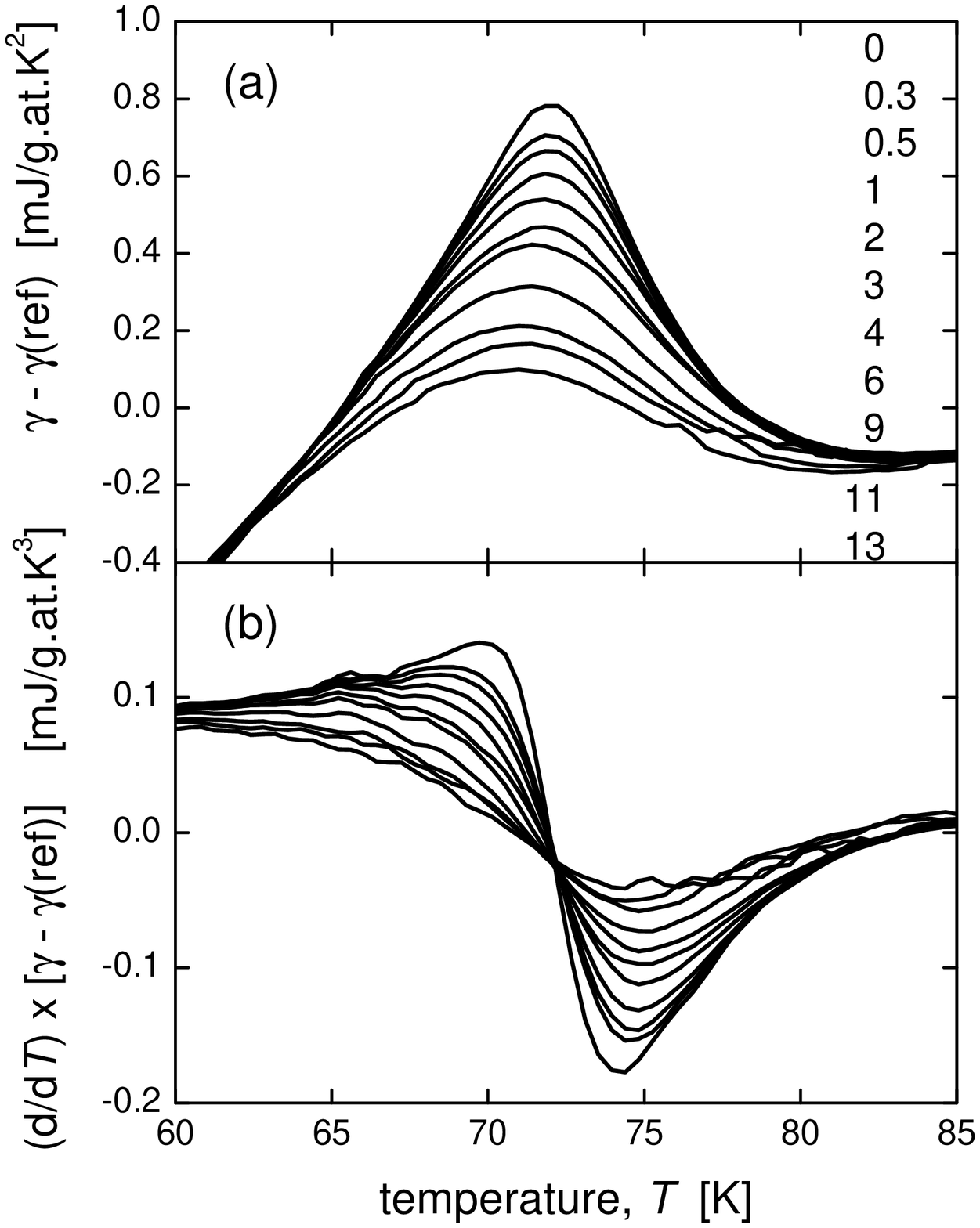}}
\caption{\small (a) The field dependence of the specific heat
coefficient $\gamma$-$\gamma$(ref) for Bi-2212. The field is shown
in units of Tesla. (b) the derivative,
${{d}\over{dT}}\times(\gamma-\gamma$(ref)) showing the
field-broadening of the inflexion points.}
\end{figure}

To gain more insight into the shape of the anomaly we have
simulated the effects of broadening by integrating sharp specific
heat transitions over gaussian distributions of $T_c$. The data
are compared with similar plots for some of our Bi-2212 data. To
cover a wide range of situations we have considered (a)
transitions with a simple MF step, (b) pure $\ln t$ fluctuations
with no MF step, and (c) admixtures of the two. In all following
cases $C$ corresponds to the difference $C-C_{normal}$.

For the simulated MF specific heat function we adopt a form which
conserves entropy at $T_c$, and has a step $\Delta C_{mf}=2$ at
$T_c$:
\begin{equation}
\begin{array}{rclr}
 C_{mf}(x) & = & x(3x^2-1)  & \text{for}\,\,x = T/T_c \leq 1,
\\ C_{mf}(x) & = & 0 & \text{for}\,\,x >1.
\end{array}
\end{equation}

For the fluctuation specific heat we assume
\begin{equation}
 C_{fluc}(x) = \ln(1/\mid x-1\mid)\,\,\,\text{for all}\,\, x
\end{equation}
This has a symmetrical divergence at $T_c$. For the distribution
of $T_c$ values we assume a Gaussian function centered on $T_{co}$
with half-width $w$ = 0.001, 0.02, 0.05, 0.1, 0.15 and 0.2:
\begin{equation}
 P(y) = {{1}\over{w\sqrt{2\pi}}} \times \exp(-{{y^2}\over{2w^2}});\,\,\,\,\,\, y=T_c/T_{co}.
\end{equation}

The specific heat resulting from this $T_c$ distribution is
\begin{equation}
 C(T) = \int_0^\infty{C(x)P(y)dy}
\end{equation}
\begin{figure}
\centerline{\includegraphics*[width=85mm]{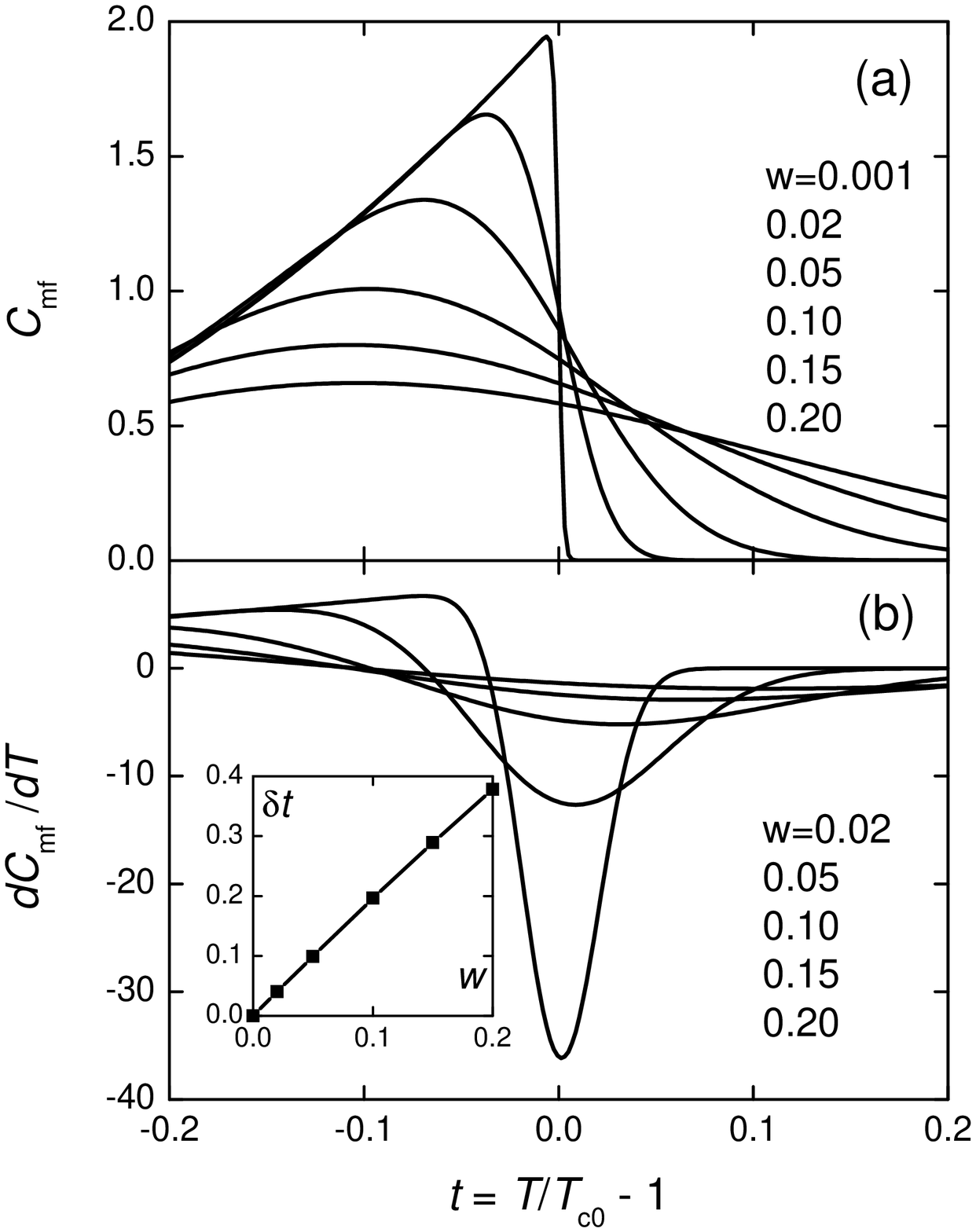}}
\caption{\small (a) The MF specific-heat anomaly modelled using
eqn. (1) with a normal distribution of $T_c$ values with half
width, $w$ = 0.001, 0.02, 0.05, 0.10, 0.15 and 0.20. (b) The
derivative d$C_{mf}$/d$T$. Inset: $\delta t$ (defined in text)
versus $w$.}
\end{figure}

{\it Results}. For a range of widths $w$, plots of C and $dC/dT$
versus $t = (T/T_{co})-1$ are shown in Fig. 3 for $C_{mf}$ and in
Fig. 4 for $C_{fluc}$ . Admixtures $C_{mix} = C_{mf} + mC_{fluc}$
with $m=$ 0, 0.2, 0.4, 0.6, 0.8 and 1 are plotted versus $t$ in
Fig. 5(a) and (b) for $w=0.02$.

It is important, firstly, to note that in spite of the very
different $T$-dependencies of the unbroadened anomalies, the
region over which broadening effects are important is very similar
for $C_{mf}$, $C_{fluc}$ and $C_{mix}$. This shows that, though
the Andersen model does not include fluctuations, the addition of
fluctuations would not alleviate the disparity between their model
and the experimental data. In particular, inclusion of
fluctuations will not result in a more narrow negative-curvature
region around $T_c$ as seen in the data. We also find that the
relative positions of key features of the $T$-derivatives of the
broadened curves are insensitive to the detailed $T$-dependencies
assumed for $C_{mf}$ and $C_{fluc}$. From these plots we find the
locations of key features and determine their relation to the
half-width $w$ of the distribution of $T_c$ values.

(i) For the MF anomaly, shown in Fig. 3, the most useful features
are the positions of the points of maximum negative and positive
curvature $d^2C/dT^2$ at $t_-$ and $t_+$ respectively, and the
difference $\delta t = t_+ - t_-$. In the inset to Fig. 3(b) we
show $\delta t$ plotted as a function of the half-width $w$. Over
most of the range of $w$, $\delta t\approx 2.0w$.

\begin{figure}
\centerline{\includegraphics*[width=85mm]{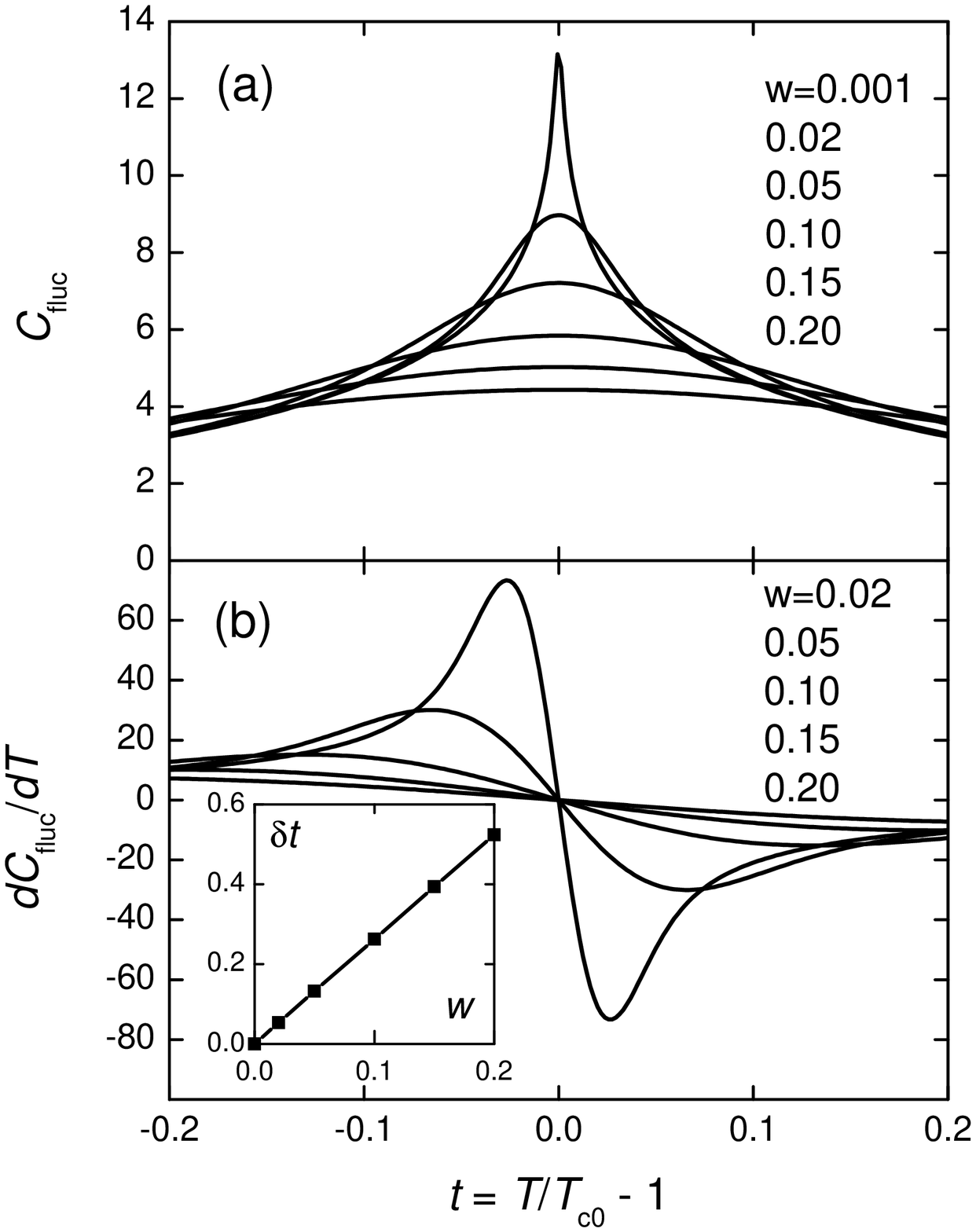}}
\caption{\small (a) The fluctuation specific-heat anomaly modelled
using eqn. (2) with a normal distribution of $T_c$ values with
half width, $w$ = 0.001, 0.02, 0.05, 0.10, 0.15 and 0.20. (b) The
derivative d$C_{fluc}$/d$T$. Inset: $\delta t$ (defined in text)
versus $w$.}
\end{figure}

(ii) For the symmetric fluctuation anomaly, shown in Fig. 4, the
most useful features are the positions of the points of maximum
positive and negative slope $dC/dT$ at $t_-$ and $t_+$
respectively, and the difference $\delta t = t_+ - t_-$.  In the
inset to Fig. 4(b) we show $\delta t$ plotted as a function of
$w$. Over the entire range, $\delta t\approx 2.63w$.

(iii) For the admixture of a MF anomaly and fluctuations, Fig. 5
shows plots of $C_{mix} = C_{mf} + mC_{fluc}$ for $w$=0.02 with
$m$ = 0, 0.2, 0.4, 0.6, 0.8 and 1. These are useful for comparison
with typical cuprate specific heat data and cover the range from
asymmetric ($m$=0) to almost symmetric ($m$=1) anomalies. A value
of $m \approx$ 0.8 is appropriate for weakly-overdoped Bi-2212.
From Fig. 5(b) we obtain the positions of the points of maximum
positive and negative slope at $t_-$ and $t_+$, respectively, and
the difference $\delta t = t_+ - t_-$. For larger values of $m$,
typical of our Bi-2212 samples, we find $\delta t \approx 2.7w$.
For $m$=$\infty$ (pure fluctuations) we found, above, $\delta t
\approx 2.63w$.

\begin{figure}
\centerline{\includegraphics*[width=85mm]{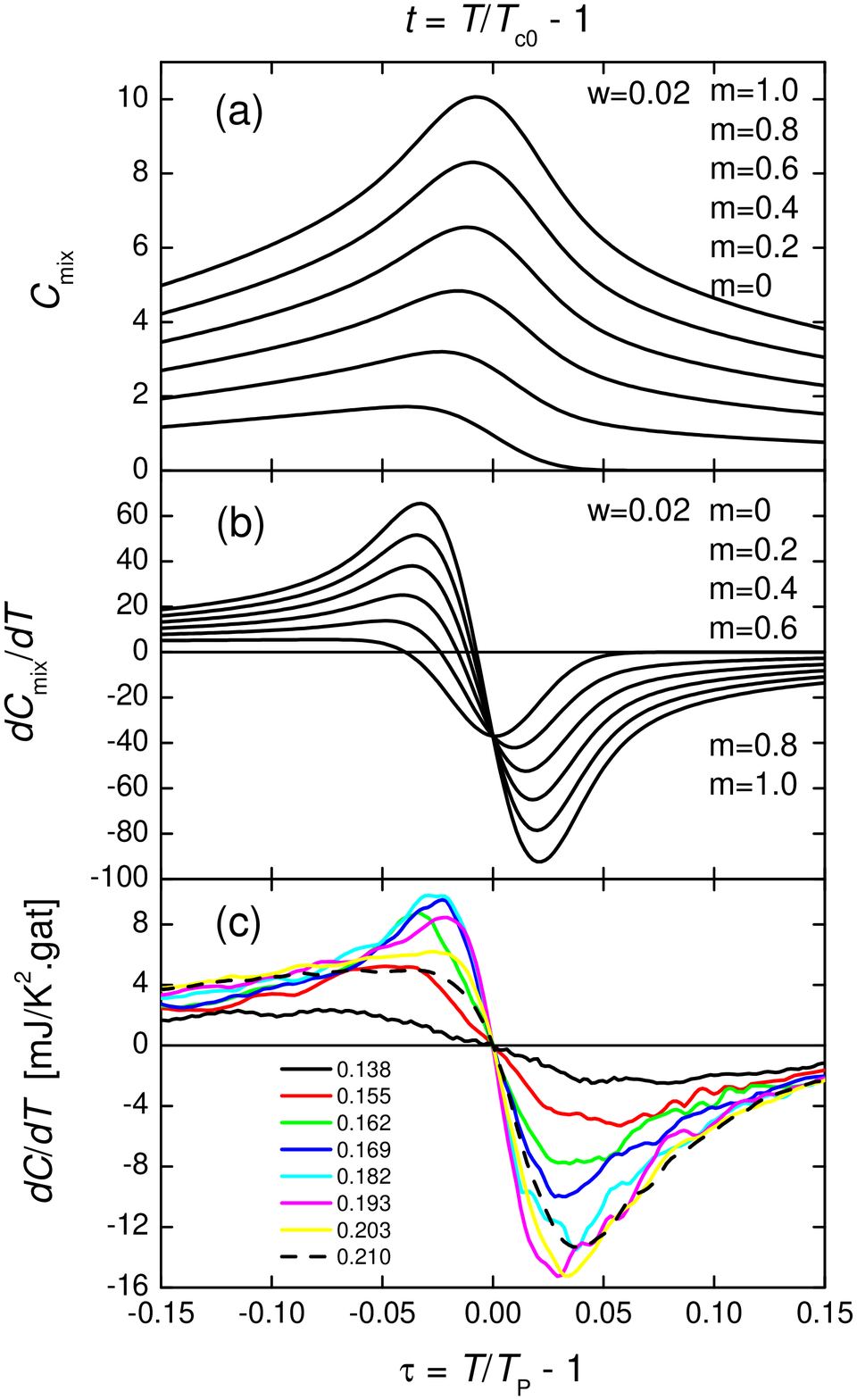}}
\caption{\small (a) The admixture specific-heat anomaly $C_{mix} =
C_{mf} + mC_{fluc}$ with half width, $w$ = 0.02 and various mixing
ratios, $m$, plotted versus $t = (T/T_{co})-1$. (b) The derivative
$dC_{mix}/dT$ vs $t$. The zero crossing occurs at $T_p$ and the
curves intersect at $T_{c0}$. (c) $dC/dT$ vs $\tau$ for Bi-2212 at
different doping levels (annotated) showing a crossover from pure
fluctuations for $p<0.162$ to an admixture with a MF step for
$p>0.162$. }
\end{figure}

Now we compare this model data with measured data for Bi-2212.
Plots of $dC/dT$ versus $\tau = T/T_p-1$ are shown in Fig. 5(c)
for eight values of $p$ (ranging from well underdoped to well
overdoped). Comparison with Fig. 5(b) shows that $T_c$ is only
1-2\% above $T_p$, so separations of peaks in $t$ and in $\tau$
are almost identical. When underdoped the anomaly is pure
fluctuations with no MF step while the increase in MF step is
evident for overdoping from the increasingly different relative
magnitudes of the positive and negative peaks in $dC/dT$.

We estimate the transition width, $w$, from the separation of the
positive and negative peaks in $dC/dT$. For all samples with $p
\geq$ 0.169 the separation is $\delta t \approx$ 0.05 - 0.06.
Taking $\delta t \approx$ 2.7$w$ for $m\approx$1 (see Fig. 5(b))
gives $w \approx$ 0.019 - 0.021 for the half width of the
distribution of $T_c$ values. For $p$=0.162 we have $\delta t
\approx$ 0.073 or $w \approx$ 0.028, and for $p$=0.138 we have
$\delta t \approx$ 0.14 or $w \approx$ 0.056. All these values are
in good agreement with the values of the half-width $\Delta t$
shown in Fig. 4 of our previous work\cite{Loram}.

As a check on our previous method\cite{Loram} for estimating these
half-widths, $\Delta t$, we show in Fig. 6 plots of $C_{mix}$
versus $\log_{10}(t)$ and $\log_{10}(t^*)$, respectively, for $w$
= 0.02 and $m$ = 0 to 1, where $t^* = \sqrt{t^2 + \Delta t^2}$.
Our normal procedure is to choose a value of $\Delta t$ that just
averts the negative curvature close to $T_c$ seen in plots of
$C_{mix}$ versus $\log_{10}(t)$, and it is evident from the solid
curves that the choice $\Delta t = w$ achieves this result. This
confirms that estimates of the broadening $\Delta t$ from plots of
$\Delta C$ versus $\log_{10}(t^*)$ give reliable values comparable
to the true half-width $w$. The plots in Fig. 6 also show that
this procedure provides a reliable estimate of the MF step $\Delta
C_{mf} \approx$ 1.5 even in the presence of large fluctuations.

\begin{figure}
\centerline{\includegraphics*[width=85mm]{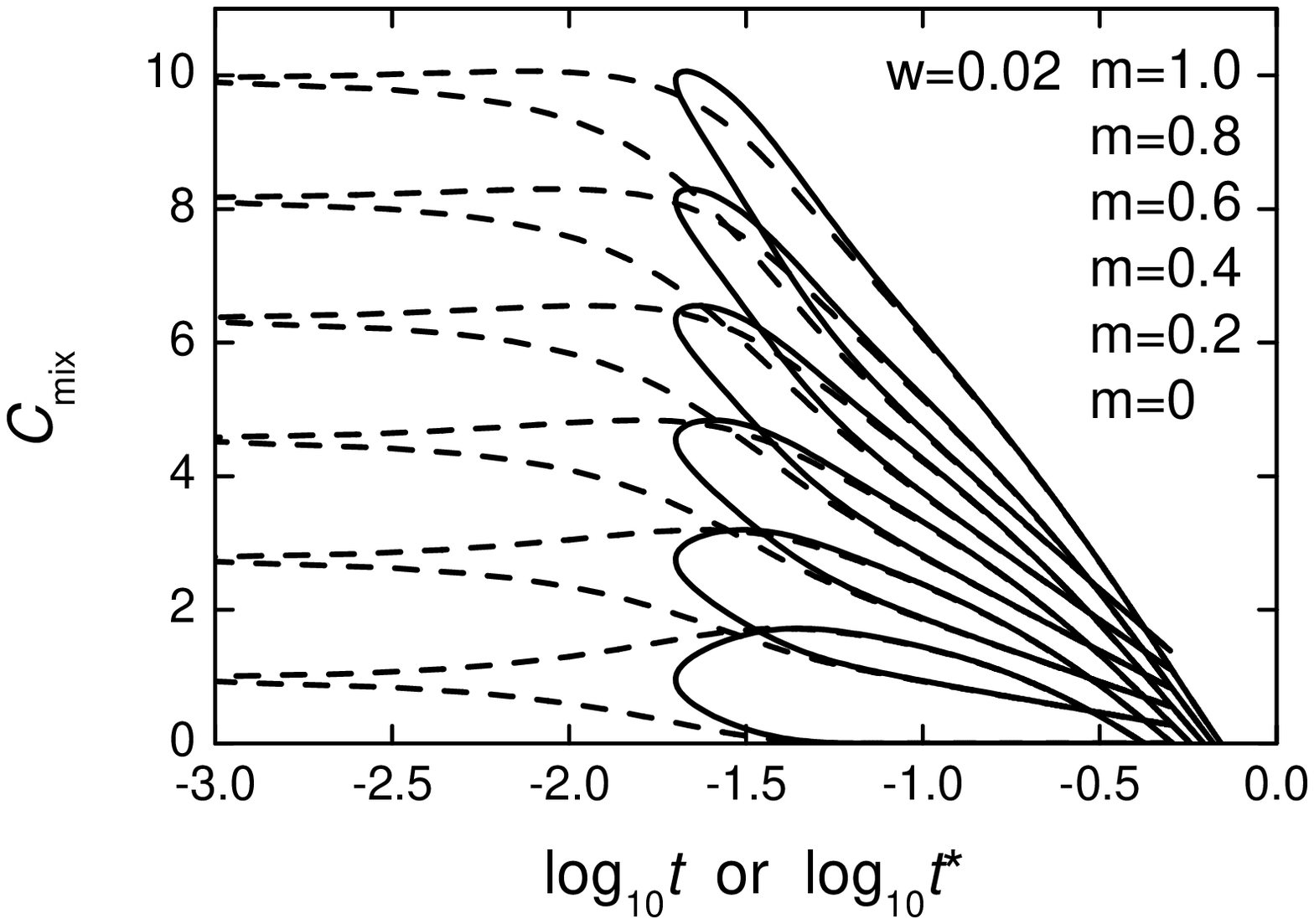}}
\caption{\small The admixture specific-heat anomaly $C_{mix} =
C_{mf} + mC_{fluc}$ with half width, $w$ = 0.02, plotted against
$\log_{10}(t)$ (dashed curves) and against $\log_{10}(t^*)$ (solid
curves) with $t^* = \sqrt{t^2 + \Delta t^2}$ and setting $\Delta t
= w$.}
\end{figure}

Turning to the MF calculation of the specific-heat anomaly by
Andersen {\it et al.}\cite{Andersen}, and recalling that their
model does not include fluctuations, we compare their results
(shown in Fig. 4(b) of their paper) with our plots for a
fluctuation-free broadened MF step, shown here in Fig. 3. From the
location of the temperatures $T_-$ and $T_+$ of their maximum
negative and positive curvature, we obtain $\delta t = (T_+ -
T_-)/T_{av} \approx$ 0.146, 0.204 and 0.232 for their curves for
$\delta g/t =$ 1.0, 1.5 and 2.0, respectively. Taking $\delta t =
2w$ from Fig. 3(b) we find $w \approx$ 0.073, 0.102 and 0.116,
respectively, as the half-widths for the three curves. This agrees
with the $\approx 20\%$ breadth of the transition that they quoted
for $\delta g/t = 1.5$, the value which best accounts for the
gap-maps. This may be compared with the similar analysis of our
Bi-2212 data from which we obtained $w \approx 0.02$ over most of
the doping range, a factor of five, or more, lower than the
Andersen estimate based on gap-maps. It surely cannot be claimed
that the spectroscopic and thermodynamic data are consistent.
Indeed, given our evidence that the spread of $T_c$ values is
rather narrow, their calculations show plainly that the
inhomogeneity in the gap-maps cannot result from gross pairing
energy disorder in the bulk.

To conclude, we have shown that the inference of pairing
inhomogeneity from STS gap maps, and the resultant transition
broadening, is inconsistent with the specific heat data which
exhibit sharp features with transition widths of the order of 3K
in Bi-2212. It is not possible with any broad spread of SC gaps to
have strong $T$-dependences over a narrow $T$-range. This concurs
with a recent analysis of quasi-particle scattering seen in
spatial modulations of STS data. The Fourier transform of these
patterns reveals spots, corresponding to scattering $q$-vectors,
that are far too narrow for the presumed broad distribution of SC
gaps\cite{Storey}. The inference of large-scale gap and pairing
inhomogeneity at ($\pi$,0) does not appear to be supported by the
wider thermodynamic\cite{Loram}, NMR\cite{Bobroff},
ARPES\cite{Kanigel} and tunneling\cite{Storey,Misra} data.

\end{document}